\documentclass{sig-alternate}
\usepackage[pdftitle={DPp is not dPP},
  setpagesize=false,
  colorlinks=true,
  citecolor=black,
  linkcolor=black,
  urlcolor=black]{hyperref} 
\usepackage{graphicx}
\graphicspath{{.}{../img/}}

\newcommand*{\textC}[2]{\textsc{#1}\raisebox{4pt}{{\tiny #2}}} 
\newcommand*{\Ca}[1]{\textC{#1}{V}} 
\newcommand*{\Cb}[1]{\textC{#1}{S}} 
\newcommand*{\Cc}[1]{\textC{#1}{C}} 
\newcommand*{\Qu}[1]{\textit{``#1''}} 
\newcommand*{\refq}[1]{#1} 

\begin{document}
\toappear{\relax}
 
\title{Distributed-Pair Programming can work well\\
  and is not just Distributed Pair-Programming}

\numberofauthors{3}
\author{
\alignauthor
Julia Schenk\\
       \affaddr{Freie Universit\"at Berlin}\\
       \affaddr{Institut f\"ur Informatik}\\
       \affaddr{14195 Berlin, Germany}\\
       \email{Julia.Schenk@inf.fu-berlin.de}
\alignauthor
Lutz Prechelt\\
       \affaddr{Freie Universit\"at Berlin}\\
       \affaddr{Institut f\"ur Informatik}\\
       \affaddr{14195 Berlin, Germany}\\
       \email{prechelt@inf.fu-berlin.de}
\alignauthor
Stephan Salinger\\
       \affaddr{Freie Universit\"at Berlin}\\
       \affaddr{Institut f\"ur Informatik}\\
       \affaddr{14195 Berlin, Germany}\\
       \email{Stephan.Salinger@inf.fu-berlin.de}
}
\maketitle
\begin{abstract}
\emph{Background:}
Distributed Pair Programming can be performed via screensharing or via a
distributed IDE.
The latter offers the freedom of concurrent editing (which may be helpful or
damaging) and has even more awareness deficits than screen sharing.
\emph{Objective:}
Characterize how competent distributed pair programmers may handle this
additional freedom and these additional awareness deficits and characterize
the impacts on the pair programming process.
\emph{Method:}
A revelatory case study, based on direct observation of a single,
highly competent distributed pair of industrial software developers during a
3-day collaboration.
We use recordings of these sessions and
conceptualize the phenomena seen.
\emph{Results:}
1.~Skilled pairs may bridge the awareness deficits
without visible obstruction of the overall process.
2.~Skilled pairs may use the additional editing freedom in a useful limited
fashion, resulting in potentially better fluency of the process than local pair
programming.
\emph{Conclusion:}
When applied skillfully in an appropriate context, distributed-pair
programming can (not will!) work at least as well as local pair programming.
\end{abstract}

\category{D.2.m}{Software Engineering}{Miscellaneous}

\terms{Human Factors}

\keywords{distributed software development, collaboration, pair programming,
  distributed pair programming}

\section{Introduction}

Kent Beck defines pair programming (PP) as follows
\emph{``Write all production programs with two people sitting at one
  machine.  
  [\ldots]
  Pair programming is a dialog between two people simultaneously
  programming (and analyzing and designing and testing) and trying to program
  better.''}
\cite[p.26]{Beck04}.

Pair programming has the obivious disadvantage of blocking two developers
rather than just one, but has many potential advantages as well:
It can be useful to reduce elapsed time, to reduce defect density,
to improve program design, to make sure more than one person is familiar with
each part of the code, to increase the amount of knowledge available when
solving a task, to increase focus and keep up discipline, to accelerate
learning, and to build within-team trust, among other things; see for instance
\cite{HanDybAri09,Nosek98,WilKes02}.
There is a substantial empirical research literature about pair programming; see
Williams' summary \cite{Williams11} for a good overview.
The evidence is mixed, but it appears that in the right circumstances pair
programming is a very helpful technique.
Unfortunately, research cannot yet characterize well which are these right
circumstances.

Kent Beck thinks that one of the factors is pair programming skill:
\emph{``Pair programming [...] [is] a subtle skill''} 
\cite[p.100]{Beck99}.
We agree and have set out since 2004 to decipher the actual work process of
pair programming in order to characterize this skill.
We have just completed the foundation \cite{SalPre13-baseconbook}
and the first results are now starting to appear
\cite{SalZiePre13-pproles,ZiePre14-ppknowtrans}.

So let us assume there are two skilled pair programmers who have decided the
circumstances are such that they should practice pair programming. 
But what if they are not at the same site?
A recent survey of 326 practitioners mostly from North America and Europe 
\cite{Williams12} had
96\% of them answer that their project 
involved distributed development (49\% even trans-continentally) -- and these
were all practitioners of agile methods!

A possible solution is to mediate the pair programming collaboration by
technological means. For synchronous collaboration, as it is the case for
distributed pair programming, computer-supported cooperative work
research (CSCW) considers two approaches of such
technological support: sharing a single-user application (screen sharing)
or using a dedicated collaboration tool (here: a distributed IDE)
\cite{Hanks02}.

For verbal communication we assume a good hands-free audio connection of the
partners, but in either case the awareness of the physical actions and
reactions of the 
partner is far lower than in local pair programming and the tool determines the
availability of workspace awareness information as well as the participants'
interaction capabilities; the tool has a
major impact on the collaboration process and can get in the way.

Synchronous remote collaboration is known to be difficult in general
(``Distance matters'', \cite{OlsOls00}), but for agile software development
with its strong emphasis on intense human interaction it sounds like
an almost paradoxical idea not likely to succeed \cite{LukSch07}.

The research presented here originally set out to characterize
the process differences between local pair programming and distributed pair
programming, but this has proven too difficult:
We study actual problem-solving of industrial software developers
and so no two observations are ever exactly comparable.
Therefore, it is hard to determine whether and where any
particular behavioral difference observed between two sessions comes from
characteristics of the pair, 
from characteristics of the tasks, or 
from the fact that one pair is working locally and the other distributedly.
We will have to set a less ambitious research goal.

\subsection{Research Approach and Objective}
\label{overallapproach}
\label{objective}
We will follow a case study approach in the sense of Yin \cite{Yin03},
each case being one pair of developers, and ask
the following research question:
\begin{quote}
  How do distributed pair programmers cope with those technology-induced
  influences on their situation that threaten successful collaboration?
\end{quote}
We will derive the set of these influences theoretically (there are only two),
formulate issues of interest regarding these,
use direct observation of distributed pair work (by means of recordings),
conceptualize the behaviors seen (by means of Grounded-Theory-style open
coding and constant comparison),
interpret the behaviors to evaluate the issues,
and then draw conclusions.

We originally studied several pairs, but eventually decided to focus on only
the most competent and successful one of them and so
turn the study from a multiple-case study delivering a complicated, mixed
message into a revelatory\footnote{Yin discusses five situations in which a
  single-case study will often be sufficient: critical case, extreme case,
  unique case, revelatory case, and representative case
  \cite[pp.39--42]{Yin03}.
  Our case can be viewed as critical, extreme, relevatory, or several
  of these. We will use elements from the extreme and relevatory perspectives
  but do not discuss the discrimination further.}, 
single-case study providing the existence proof
for something that has not been described before:
Industrial distributed pair programming that appears to be as effective as if
it had been local pair programming.
Conventional wisdom in the field of CSCW suggests 
such a thing does not happen \cite{OlsOls00}.

We will now derive the above-mentioned influences in Subsections
\ref{awareness} and \ref{editingfreedom} and underway
discuss the differences between the three basic types of tool
support (RPP, DPP, and eDPP; we will study eDPP only) 
in Subsections \ref{RPP} and \ref{DPP}.

\subsection{Influence 1: Reduced Awareness}
\label{awareness}

In local pair programming a rich set of cues is available from posture,
gesture, touching the partner's body, handling other objects, gaze direction,
facial expression, and all kinds of vocal and sub-vocal noises.
These cues provide lots of information relevant to the interaction of the
pair members and contribute to the pair members' awareness of their partner
and the joint work.
In a distributed setting, the cues are reduced immensely:
the vocal noises are transmitted much less finely and the other cues are
essentially no longer available at all.\footnote{At the current state of
  technology, pairs tend to prefer working without a video connection
  \cite{Hanks02,StoWil02}.
  This was also true of our participants, who had a video connection
  available but turned it off. 
  They stated that (1)~the value it added was very low and 
  (2)~the movement in the video image 
  (on the screen rather than at the edge of one's field of vision)
  distracted from the actual work.}
This reduction of physical awareness is a major reason why synchronous
collaboration is considered inferior when it has to be mediated by
technological means \cite{GutGreRos96}.

Another reduction in awareness concerns the state of and the activity in the
joint computerized workspace (workspace awareness).
To explain the workspace awareness situation of our study, we need to delve
into the differences of tools for RPP-style collaboration (which we do not
study) and DPP-style collaboration (which we do study).

\subsection{Remote Pair Programming (RPP)}
\label{RPP}
The collaboration mode to which we refer as remote pair programming (RPP)
is usually implemented by screen sharing.\footnote{There are also RPP pairs using
  a shared text terminal, see \url{http://www.pairprogramwith.me}.
  We ignore these here.}
With respect to viewing,
one pair member's desktop (the local side) is transfered to the partner's
computer (the remote side) via software such as 
VNC, Adobe Connect, Skype, or many others. 
This provides strict WYSIWIS (What You See Is What I See) \cite{SteBobFos87}
and therefore the awareness of the technical workspace is good: 
If the remote screen is at least as large, both
partners see the same screen content, just as in local pair programming.
Depending on the software used, the mouse cursor may be an exception.
  
With respect to operating the computer, either all input device streams are
merged into a single input stream or, more commonly, 
strict floor control is used which will allow input only by one participant
at any time \cite{SteBobFos87}. 

RPP involves an asymmetric situation:
The remote partner is disadvantaged when operating the computer because she
suffers from twice the network latency, possibly from bandwith limitations,
and from having to cope with the partner's IDE configuration (e.g. colors,
layout, key bindings, and so on).
The latter is true for local pair programming as well but the former tends to
make RPP less than ideal.


\subsection{Distributed Pair Programming (DPP and eDPP)}
\label{DPP}
Distributed pair programming (DPP) removes the asymmetry of RPP.
DPP uses a dedicated collaboration tool: a distributed IDE.
Such a tool replicates the files to both participants and makes all editing
operations local, transfering them to the other side to keep the files in sync.
The tool will also transfer the cursor position and all text editor view changes
(scrolling, jumping, file switching).
This approach makes network latency easily bearable, makes bandwith
requirements small, and
each participant can use his own IDE settings and configuration. 

Some such tools (such as Sangam \cite{HoRahGeh04} or XPairtise \cite{SchLuk09})
enforce a strict pair programming mode: 
Only one partner is allowed input at any time and the views are kept in sync
as far as viewport layout permits, resulting in approximate WYSIWIS.
This article's title calls this work mode distributed pair-programming.

Other tools\footnote{in particular
  VSAnywhere (\url{https://vsanywhere.com}) and 
  Saros \url{http://www.saros-project.org/} (see Section~\ref{saros}),
  as well as a few server-based IDEs specialized for developing web
  applications and used via a web browser, 
  notably Codenvy (\url{http://codenvy.com}).}
allow editing freedom, that is, independent and concurrent editing and viewing:
There is a ``follow mode'' (``Shared Browsing'' in \cite{SchLuk09})
in which pair member A's view follows the
activity of the partner B, just like in strict DPP,
but A can leave it and become active at any time to view other
files, other regions of the same file, or to make changes, even in the
exact same spot that B is currently working on.
Operational transformation \cite{SunEll98} is used to ensure 
unique and consistent editing results despite the race conditions.
If this modality is used occasionally but the pair can still be said to
perform pair programming, then we call the resulting work mode extended
distributed pair programming (eDPP) or, as the title calls it, 
distributed-pair programming.

In terms of workspace awareness, DPP is much more problematic than RPP:
(1)~The number of lines and characters in the editor window may be different,
so that one partner may see more source code content than the other, which may
occasionally lead to communication problems.
Markup color differences may create similar issues.
(2)~When using other applications such as a web browser or when testing
the application, their views are not transmitted to the partner.
Switching to screen sharing is required to bridge these situations
where needed.
(3)~Even within the IDE, not all views may be synchronized, in particular the
package explorer view and the various pop-up dialogs; this is tool-dependent.

If the pair makes use of editing freedom in eDPP, the awareness situation
becomes worse: 
After leaving follow mode, the former observer may no longer see all of what
the former driver is doing and the former driver may not even become aware of
this fact.

\subsection{Influence 2: Editing Freedom}
\label{editingfreedom}

An eDPP tool supports the side-by-side programming work mode 
\cite[Section 3.T8]{Cockburn04}: 
The partners work mostly separately on separate but related sub-tasks
and collaborate synchronously only when and where needed
\cite{PreStaSal09-sbstype};
they may switch between pair programming and solo programming many
times. 

This is clearly no longer pair programming, so if pair programming was intended,
slipping into side-by-side programming may be problematic.
Unfortunately, once a pair programming pair makes any use of editing freedom at
all, there is no clear line indicating where pair programming ends and
side-by-side programming begins, so the editing freedom of eDPP may prove to be
a slippery slope and may threaten the original goals (such as
broadening the pair's common ground \cite{OlsOls00}) of their pair programming
session.

\subsection{Structure of this Article}

We will now sketch the method of our work
(Section~\ref{method})
before we introduce the specifics of our study setup 
including the case (Section~\ref{case}),
the eDPP tool used (Section~\ref{saros}),
expectations suggested by related work (Section~\ref{relatedwork}),
and the refined research questions resulting from all this 
(Section~\ref{refinedquestions}).
We then present the results and discussion
(Sections \ref{resultsroles} to \ref{resultsediting}: roles,
awareness-related phenomena, editing-related phenomena).
After a discussion of the limitations of our work
(Section~\ref{limitations}),
we draw conclusions (Section~\ref{conclusion}).

\section{Method}
\label{method}

Here we describe how we collected the data of our study
(Subsection~\ref{datacollection}),
how we analyzed the data (\ref{dataanalysis}),
and in what format we present the analysis results
(\ref{notation}).

\subsection{Data Collection Method}
\label{datacollection}

We have been collecting recordings of local pair programming sessions since 
2004 and of distributed pair programming sessions (most of them
eDPP, but also some RPP) since 2010;
some of them come from lab settings with student programmers, 
but most of them (37 sessions of 25 different pairs
involving 38 persons from 8 different companies from various industries) 
involve professional
software developers doing their normal development work. 
The sessions have a typical length of one to three hours.

For recording an eDPP session, we invite each of the two partners
into a separate(!) Adobe Connect web conferencing session with us as the other
end, have them each share their screen (including webcam and audio), place the
two web browsers showing these web conference sessions on a single large
portrait-view monitor and record its screen with Camtasia Studio.
The recording contains for each pair member all desktop activity, audio, and a
small webcam view as seen from atop that member's monitor.

Note that the pair itself lacks such comprehensive overview:
Each of the two hears the partner's audio (via Skype, using headsets) 
but sees only his own desktop and IDE;
they chose not to use video because they found it more distracting than
useful, which is in line with claims elsewhere \cite{Hanks02,StoWil02}.
The pair volunteered to be recorded for research purposes after they had
received support in setting up Saros and general advice on DPP from us when
they had become interested in using Saros.

\subsection{Data Analysis Method}
\label{dataanalysis}

Our data analysis primarily uses elements of the Grounded Theory Methodology
(GTM, \cite{StrCor90}) but 
does not aim at producing an actual Grounded Theory
and does not claim to use GTM in full.
We mainly use open coding \cite[Section~II.5]{StrCor90}
and constant comparison \cite[Section~II.1]{StrCor90}, 
with the theoretical sensitivity \cite[Section I.3]{StrCor90}
oriented as described in Section~\ref{refinedquestions} and 
an overall workflow as described in Section~\ref{overallapproach}.

\subsection{Notation}
\label{notation}

We will report our results narratively and join it with the discussion and
interpretation in order to avoid redundancy.
Our results are primarily individual concepts arising from
open coding and so may become hard to identify within the
narrative. 
Therefore, whenever we report one of the resulting concepts, we
typeset its name in small caps to make it visible.

Furthermore, the concepts have been elaborated to different levels of
precision and accuracy.
Since there is not enough space to present the concepts in full detail, we 
discriminate three rough levels of elaboratedness as follows.
``V'' (as in \Ca{Some Concept}) represents vague, informal concepts that
appeal to
intuition and for which we have produced hardly more definition than their
name.
``S'' (\Cb{Some Concept}) marks semi-complete concepts
for which a concrete definition is available but where we expect that
definition to be incomplete and/or unstable (from the point of view of
continued research on the topic).
``C'' (\Cc{Some Concept}) marks completely elaborated concepts that we
consider stable.

\section{Context and Refined Questions}
\label{context}

This section describes the specific context of our case study:
The pair and what they did (Subsection~\ref{case}) and
the specific eDPP tool they used (Saros, Subsection~\ref{saros}).
From these, we then derive finer-grained research questions 
(Subsection~\ref{refinedquestions}).

\subsection{The Case}
\label{case}

Our pair consists of two male German software developers, 
J1 working at a software development service company's home site and 
J2 working at its customer in a different city.
J2 is an intermediate-level software developer and has been working on the
customer's information systems and workflow automation software in the radio
broadcasting domain for a long time.
We will call him ``Dev'' in this article.
J1 is a senior developer and software architect who is now assigned 
the task to perform an overall review and design optimization
of J2's workflow automation software together with J2.
We will call him ``Arch'' in this article.
The two work together on this for several days;
they had other stretches of collaboration previously
and so are well-familiar with each other.
The recordings we made of this and have used for this article cover 
three days of this collaboration; 
there are 7 recordings (JA2\footnote{We use global IDs for all our
  recordings; JA1 was a Saros trial two weeks before.}
to JA8) between 0:42 hours and 2:01 hours length 
and one (JA9) of 5:26 hours.

The first sessions focus on transfering domain knowledge from Dev to Arch,
performing a joint design review and discussion, 
and transfering design knowledge from Arch to Dev.
They also involve refactorings.
Later sessions revolve around the re-design and re-implementation of one
complete module and most of the observations
conceptualized below stem from these.

\subsection{The Tool: Saros}
\label{saros}
Saros \cite{SalOezBee10-saros}\footnote{http://www.saros-project.org} is an
open source Eclipse plugin for eDPP that realizes
a distributed editor within the Eclipse IDE which supports up to five
users.
We have been developing Saros since 2006, initially focussing on
functionality and recently doing mostly architecture consolidation and
usability work. 
Saros has now reached industrial strength and marks the state of the art in
its area, on par with VSanywhere and clearly superior to all other DPP and
eDPP tools we are aware of.
In particular in terms of the workspace awareness framework of Gutwin and
Greenberg \cite{GutGreRos96}, Saros provides all workspace awareness
information relevant for real-time collaboration.

For describing Saros, we will make use of the patterns for computer-mediated
interaction by Lukosch and Sch\"umer \cite{LukSch07,SchLuk09} and Capitalize
The Respective Terms.

In Saros, one participant sends an Invitation for a Collaborative Session to
another 
participant who will automatically receive a local copy of the files of one or
more shared Eclipse projects.
Saros lets them concurrently view or edit the same or different files in the
Shared Editor.
It uses social rather than technical Floor Control (alongside
with Operational Transformation to keep all copies in sync) and Conflict
Detection to safeguard against external modifications.
There is also a simple Shared Graphical Editor for sketching.
Activities in other views such as the package explorer or popup dialogs are
not transfered to the partner.
A combined User List and User Gallery shows all available users with their
Availability Status as well as the current session participants, their markup
color and the current usage of follow mode.
The file annotations and editor annotations shown in Figure~\ref{fig:saros}
provide awareness of the partner's actions and location in the shared files:

\begin{figure*}
\centerline{\includegraphics[width=15cm]{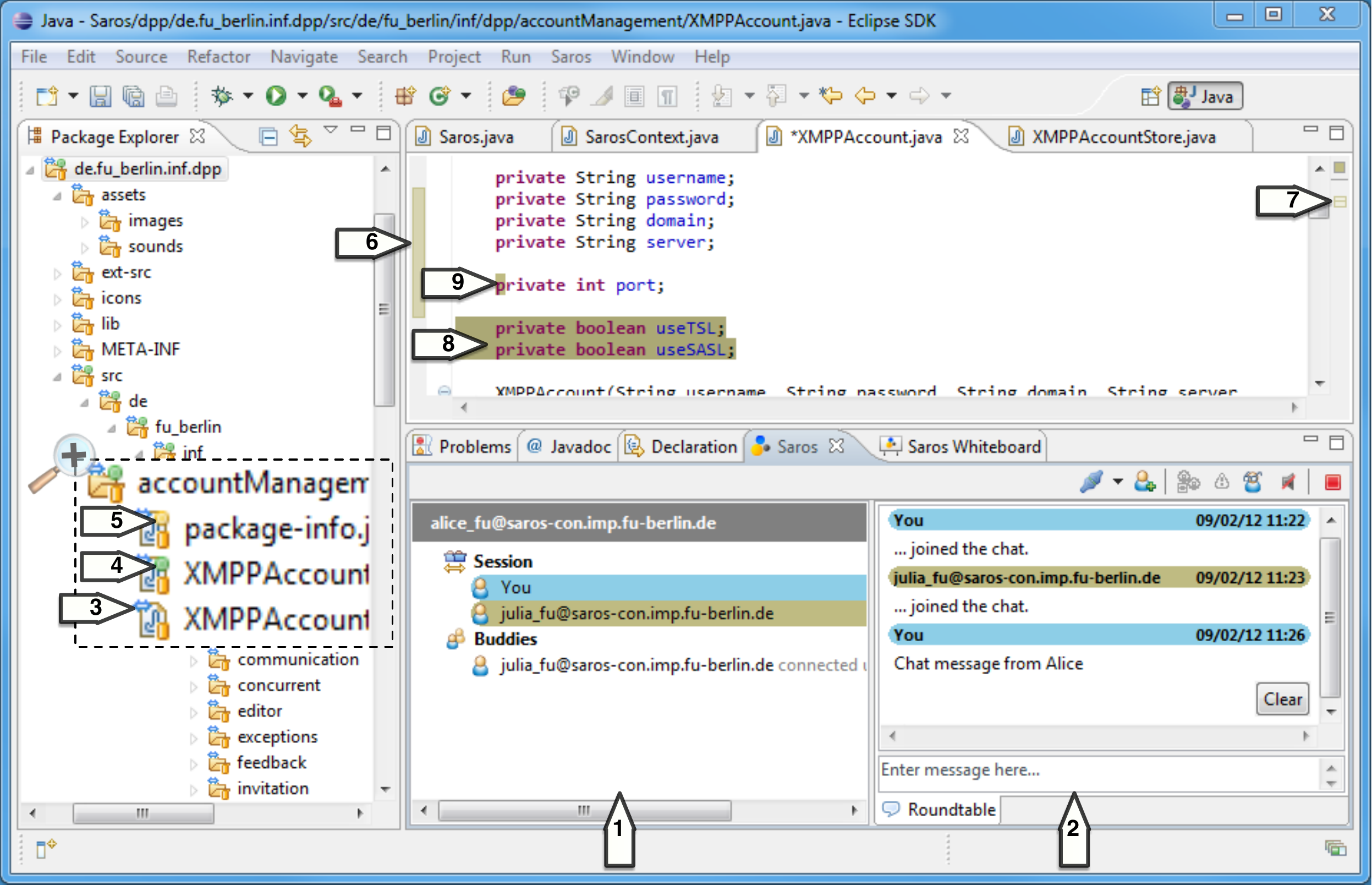}}
\caption{Screenshot of an ongoing Saros eDPP pair programming session.
  The awareness markup is highlighted; see the text for explanation.}
\label{fig:saros}
\end{figure*}

\noindent
1~The main Saros view shows the users, their colors, and the follow mode
indicator (not shown).\\
2~A session-independent Embedded Chat can be used for coordination before and
written communication during the session.\\
3~A double arrow marks the files shared in a session.\\
4~A green dot marks files currently being viewed.\\
5~A yellow dot marks files currently open.\\
6~A Remote Field of Vision annotation indicates the line range seen by the
partner if viewports overlap.\\
7~A Remote Field of Vision annotation (pseudo-scrollbar) indicates the line
range seen by the partner even if viewports do not overlap.\\
8~A Remote Selection shows the partner's current selection; the partner's most
recent edits are highlighted in a similar color.\\
9~A Remote Cursor shows the partner's cursor position.

\subsection{Related Work}
\label{relatedwork}

There is a wide spectrum of related work that is relevant for our study
but little of it is directly connected to our research question. 
We will cluster the relevant literature and shortly discuss the
relevance of each cluster for our research question.

There is substantial empirical literature on pair programming
\cite{Williams11}, but almost all of it is quantitative and has little to offer
in terms of detailed process characterization. Of the few more qualitative
works, many use student programmers in lab or homework situations and so not
only reflect low skill levels but also contexts not realistic for industrial
use. 

Regarding global software development (GSD, \cite{ICGSE12}),
most of today's research concerns the team,
project, or organizational level and so far focuses on the problems of GSD
\cite{GomMar12}, rather than on solutions for them.
Topics are for instance communication \cite{LayWilDam06,NguWolDam08},
coordination \cite{EspSlaKra07}, and trust
\cite{Eckstein10,EspSlaKra07a,SarAhuSup11}.
Little is said about real-time collaboration of individuals and the immediate
programming process we are concerned with here.

The substantial literature on distributed work from the fields of
organizational psychology and CSCW suggests
that the awareness issues in eDPP may be extremely detrimental.
For instance, Olson and Olson's summary article ``Distance matters''
\cite{OlsOls00} describes 
the notions of technology readiness, collaboration readiness, and common
ground as important preconditions for successful collaboration.
In our setting, only collaboration readiness can (for our particular
pair!) be assumed.
Whether current eDPP technology can provide sufficient awareness and whether
it will 
allow to create and maintain sufficient common ground (on a task-specific
micro-level), in particular during uses of editing freedom, is unclear.
From this point of view, we may expect the collaboration 
to be hampered by many difficulties and 
to be much more difficult than in local pair programming.

Finally, there is the notion of
a driver role and an observer (or navigator) role in pair programming;
they are relevant for the use of editing freedom.
These roles are not mentioned by Beck at all;
their most popular source appears to be a definition of pair programming by
Williams et al. that includes the following:
\emph{``One partner, the driver, controls the pencil, mouse, or keyboard and
  writes the code. The other partner continuously and actively observes the
  driver's work, watching for defects, thinking of alternatives, looking up
  resources, and considering strategic implications. The partners
  deliberately switch roles periodically.''} \cite{WilKesCun00}
(a different version of this definition is found in \cite[p.3]{WilKes02}).

Of the very few qualitative studies done so far of pair programming, two
conclude that this driver/observer distinction is misleading and
inappropriate.

First, Bryant et al., after systematic, quantitative, and very focused verbal
protocol analysis of on-site, everyday, industrial pair programming, conclude
\emph{``in contradiction to what has previously been suggested [...] 
  the pair programmers in the sessions observed did not show a general 
  difference in  the level of abstraction of their discussions according to 
  role.''}
and in particular find that the detection of minor mistakes is done as much
by the driver as by the observer \cite{BryRomBou08}.

Second, Chong and Hurlbutt \cite{ChoHur07}, 
after informal data analysis of on-site, everyday,
industrial pair programming, formulate even more strongly
\emph{``Aside from the task of typing, we found no consistent division of
  labor between the 'driver' and the 'navigator'.''}.
They also note that
\emph{``control of the machine input had a consistent, albeit subtle,
  influence on pair interactions: the programmer that controlled machine 
  input had a distinct advantage with respect to decision-making.''}.
They assume this to be a disadvantage of local pair programming and 
recommend the use of dual keyboards and dual mice to make
driver changes maximally fluent in the local case.
For the distributed case, they advise against tools of the RPP and strict DPP
kind that enforce and hence emphasize the driver/non-driver
distinction.
Their article does not discuss the problems that may result from having and
using editing freedom in a reduced-awareness situation, however.

Both of these sources suggest that eDPP may be quite different from standard
pair programming and it remains unclear whether this will be an advantage, a
disadvantage, or can be both.

\subsection{Refined Research Questions}
\label{refinedquestions}

Based on the general discussion of the influences in Subsections
\ref{awareness} to \ref{editingfreedom},
the more specific description of our particular research setup in
Subsections \ref{case} and \ref{saros}, 
and the expectations that can be derived from related work as discussed
just above,
we now formulate several
finer-grained research questions in order to focus our attention during the
data analysis.
Two of these concern the effects of the awareness issues, 
the other four concern the use and effects of editing freedom:\\
\textbf{\refq{APhys}} (which mnemonically stands for ``awareness physical''):
When and how does the reduced physical awareness influence eDPP?\\
\textbf{\refq{AWorksp}} (``awareness workspace''):
When and how do the limitations of Saros' workspace awareness influence eDPP?\\
\textbf{\refq{FView}} (``freedom viewing''):
When and how does the pair make use of concurrent independent viewing?\\
\textbf{\refq{FEdit}} (``freedom editing''):
When and how does the pair make use of concurrent independent editing?\\
\textbf{\refq{FPos}} (``freedom positive effects''):
When and how does making use of editing freedom appear to help the process?\\
\textbf{\refq{FNeg}} (``freedom negative effects''):
When and how does making use of editing freedom appear to hurt the process?

\section{Results: Roles}
\label{resultsroles}
To provide a context for what we have observed concerning roles in DPP we first
introduce the role definitions in PP. 

For PP in general, we fully agree with the results of Chong and Hurlbutt
and of Bryant et al. that the driver/observer roles contribute little to
understanding PP properly; we have in fact started work to generate a more
meaningful set of roles \cite{SalZiePre13-pproles}.
For understanding the differences between PP and DPP, however, 
{driver} and {observer} are helpful notions.

In first approximation, we define \Cb{Driver} to mean a person operating the
computer to solve the overall task and \Cb{Observer} to mean a person in a
supporting role influencing the driver in doing this.
Even this supposedly trivial definition can be surprisingly ambiguous due to
keyboard/mouse splits, enforced typing, or in dual-keyboard/dual-mouse
settings.
In an eDPP setting, this gets worse: 
There is not just a second keyboard, but also a second cursor and screen.

What we have observed is that the \Cb{Observer} in eDPP gradually gets active
and performs activities that he could not carry out in local PP,
so we introduce the additional role of \Cb{Active Observer} for someone who 
mainly ``observes'' but also occasionally operates the computer to provide
support.

If the pair symmetrically pursues two subtasks in parallel, both
participants are \Cb{Driver}. In a discussion-only phase where computer operation is
neither required nor imminent, both participants become \Cb{Discussant} and
there is neither a \Cb{Driver} nor an \Cb{Observer}.

We shortly come back to \Cb{Discussant} 
in Section~\ref{resultsawareness} on awareness;
\Cb{Active Observer} is the main topic 
in Section~\ref{resultsediting} on concurrent editing.

\section{Results: Awareness}
\label{resultsawareness}

We cover the awareness issues by working through a number of contexts in
which they are dominated by different needs and constraints.

\subsection{Awareness in Interactive Testing}

In local PP and RPP, the awareness difference between editing text files on
the one hand and running programs for testing\footnote{or other non-editing
  actions such as consulting a web browser or email client; we do not
  elaborate these here.} on the other hand is minor.

For DPP, Lukosch and Sch\"umer \cite{LukSch07} suggest the Distributed Commands
pattern as a technical solution to support the testing activity in DPP, meaning
that if one partner runs a test command this is executed on the other's side,
too.
However, Saros provides no support for this, resulting in a \Cb{Blind Period} 
for the partner.
Interestingly, our pair indeed emulated the Distributed Commands
functionality manually: 
They performed \Cc{Parallel Testing}, where each of the two started the
test locally and they used \Cc{Verbalization} to perform identical input
actions.\footnote{We have seen the use of \Cc{Screen Sharing} in another pair
  and a third conceivable strategy (which we have not yet seen) would be 
  one-sided testing supported with only \Cc{Verbalization}.}

This behavior pertains to \refq{AWorksp}.
It appears cumbersome but we saw no indications that it (or the nature of
the resulting testing process, which for our pair used to be straightforward)
inhibited the
process beyond the few seconds of additional time required.

\subsection{Awareness of Eclipse Dialog Windows}
\label{awarenessdialogs}
When the \Cb{Driver} opens an IDE dialog (such as a wizard, search dialog, or
refactoring dialog) and the partner is in follow mode, from the awareness point
of view an ideal eDPP tool would automatically transmit all relevant
information to the remote side (e.g. via a small window-screen sharing view).
Saros does not do that yet and we expected this would be a substantial
disadvantage.

It turned out, however, that it is actually almost no problem at all.
Again, the pair used two different strategies:\\
(1)~\Cb{Awareness Bridging}: 
While operating the dialog the \Cb{Driver}
does think aloud concerning relevant dialog options and his inputs.
Note this is more than mere \Cc{Verbalization} of actions.
It occured for simple dialogs such as the ``rename'' refactoring
dialog as well as for more complex ones, e.g. the create class wizard or
the search dialog.\\
(2)~\Cc{Technology-triggered Solo}: For complex dialogs we often observed that
the unilateral operation of a dialog was preceded by an explicit negotiation
and agreement regarding the intended work step.
The \Cb{Driver} would then accomplish this step alone while the
partner simply waits for the result, for example the new class, to
appear.
We have never seen \Cc{Screen Sharing} for dialog windows.

Example for \Cc{Technology-triggered Solo} combined with \Cb{Awareness Bridging}
using the ``new package'' dialog:\\
Dev [selects method `getLatestRemoteRawFile]:
  \Qu{OK, now we should write a test case.}\\
Arch:  \Qu{Yes.}\\
Dev:  \Qu{That tests just this function}\\
Arch: \Qu{Yes.}\\
Dev: \Qu{You are with me, right?}\\
Arch: \Qu{Yes.}\\
Dev: \Qu{Then we create a test.} 
  [goes to package explorer]
  \Qu{Oh, that's missing.}
  [opens ``New Source Folder'' dialog and mumbles:] 
  \Qu{source folder} 
  [types: ``test'' in input field for folder name and mumbles:] 
  \Qu{test} 
  [hits enter to create the folder, then opens new package dialog and mumbles:]
  \Qu{package} 
  [mumbles as he types into the ``Name'' input field:]
  \Qu{Transcoder was how we wanted to call it or how did we want to call
    it? So I create the test package right.}\\
Arch: \Qu{Eeeer, we wrote it up somehwhere.}\\
Dev: \Qu{Oh yes, got it, the TODO down here: newstranscoding} \\
Dev [verbalizes typing]: \Qu{news-trans-co-ding}\\

These behaviors pertain to \refq{AWorksp} and the finding is again pleasant:
The pair appeared to be sufficiently familiar with all dialogs it used that
it was never a problem for the remote partner to not be able to see the
actual dialog.
Even better, we observed that \Cc{Technology-triggered Solo} allows the
\Cb{Observer} to relax for a moment, apparently without much risk of overtaxing
the \Cb{Driver}.

One might think the loss of the review effect during the
\Cc{Technology-triggered Solo} will be problematic, but we never found this
to be the case for our pair:
After the \Cc{Technology-triggered Solo}, the relevant result is visible and
the \Cb{Observer} inspects it.
For example, we saw the \Cb{Observer} instantly recognize a misspelled class
name when the class newly created by the partner appeared.

\subsection{Awareness During General Editing}
When editing and viewing files in normal pair programming manner, our pair
regularly enabled Saros' follow mode.
Combined with the Remote Field of Vision annotation (marked 6 in
Figure~\ref{fig:saros}), this apparently served its purpose well:
We did not notice any detours or misunderstandings that appeared to arise due
to a lack of workspace awareness; another pleasant \refq{AWorksp} result.

As for \refq{APhys}, the same holds for the lack of cues from facial
expression, gesture (nodding, shake of the head) and posture:
The pair used \Cc{Verbalization} with no apparent effort, 
and used Remote Selection or the Remote Cursor as an extended index finger.
This worked well: the pair routinely used deictic references (such as
pronouns) in their verbal communication and we observed no misunderstandings
or non-understandings that appeared to be technically induced.
For instance in the previous example, Dev talks
about ``this method'' while the referred-to method is selected.

\subsection{Awareness During Pure Discussion}

When the pair needs to decide on a design approach or a work strategy,
there are often periods during which the computer is not operated
at all, not even viewed; the pair performs dialog only and both are in the
\Cb{Discussant} role. 
This pertains to \refq{APhys}.

One might expect that physical awareness would play a pronounced role during
such times.
For the pair we observed, however, this does not appear to be the case.
Dev and Arch both appear to be happy with reclining in their chair, 
staring into 
nothingness (the ceiling, out the window, etc.) and focussing on their sense of
hearing only.
We conjecture that at least for pairs such as ours, who are well familiar
with each other, proximity and visual contact play only a minor role, if any,
for discussion phases.

\section{Results: Concurrent and Interleaved Editing}
\label{resultsediting}

The phenomena around the fact that eDPP provides editing freedom, that is, 
capabilities for
concurrent-independent viewing and editing appear to be governed by two main
factors: 
(1)~The \Cb{Activity Level} of each participant and
(2)~the \Ca{Mental Coupling} between the participants.

\Cb{Activity Level} means physical activity insofar as it contributes to
the PP process (speaking, pointing, typing, etc.).
\Ca{Mental Coupling} means the degree to which the participants follow only
one single, joint \Ca{Train of Thought} as opposed to two separate strands.
It is difficult to observe, but conceptually clear enough for our purposes.

The classical driver/observer role description implies that the observer's 
\Cb{Activity Level} is usually low.

In that view, concurrent editing is an additional mode the \Cb{Observer} can
use to increase his \Cb{Activity Level}, becoming
either an \Cb{Active Observer} (with high \Ca{Mental Coupling}) 
or a second \Cb{Driver} (with low \Ca{Mental Coupling}).
A second \Cb{Driver} means the work mode is side-by-side programming, no
longer pair programming, although there is obviously a gray area in
between.
Our pair has not used two \Cb{Driver}s; we hence only
describe the \Cb{Active Observer} case.

All concurrent and
interleaved activities run the risk of reduced \Ca{Mental Coupling} 
as well as a \Cb{Loss of Review Effect} and \Cb{Loss of Knowledge Transfer}.
Depending on the session's goal, these risks may be more or less
problematic.

\subsection{Dimensions of Concurrent Editing Activity}
Besides \Cb{Activity Level} and \Ca{Mental Coupling},
we will use five more dimensions for
discussing concurrent editing phenomena:

\begin{itemize}
\item
  \Cc{Announcement Style} describes
  whether the \Cb{Active Observer} explicitly announces his subsequent
  activity 
  (\Cb{explicit}, e.g. \Qu{Oh, wait, may I just\ldots} ),
  makes some utterance that hints at the activity
  (\Cb{implicit}, e.g. \Qu{Parens are missing}),
  or simply starts acting rightaway
  (\Cc{silent}).
  The latter avoids interrupting the \Cb{Driver}'s thinking,
  but could also make the actual activity even more disturbing
  due to a higher subjective loss of control for the \Cb{Driver}.
\item
  \Cc{Visual Coupling} describes
  whether all of the \Cb{Active Observer}'s editing happens
  on the \Cc{Same Line},
  in the \Ca{Same Block} (or same small method),
  in the \Cc{Same Viewport},
  in a \Cb{Different Viewport},
  or
  in a \Cc{Different Artifact}.
  It determines the likely amount of the \Cb{Driver}'s awareness.
\item
  \Cb{Effect on What} describes whether the \Cb{Driver} appears to change his
  plans
  due to the intervention (\Cb{Change Triggered}) or not (\Cb{Stable}).
\item
  \Cc{Effect on When}: Whether the \Cb{Driver} slows down (or stops
  completely) during the
  \Cb{Active Observer}'s activity (\Cb{Temporizing}) or not (\Cb{Unperturbed}).
  Slowing is common and fully or partially turns concurrent
  activity into merely interleaved activity.
\item
  \Cc{TypeAOA} (type of the \Cb{Active Observer}'s activity):
  One of
  \Cc{Direct Fix},
  \Cb{Contribution},
  \Cb{Addition},
  \Cb{Local Solution}.
  We will structure the remaining discusssion along this dimension and
  handle each type in a separate subsection. 
\end{itemize}

\subsection{Type ``Direct Fix''}
We observed the \Cb{Active Observer} to make small improvements in the code
just on his own instead of asking the \Cb{Driver} to do it: A \Cc{Direct Fix}.
When the \Cb{Observer} noticed a small issue in the code, 
he often decided to simply fix
the problem himself (his \Cb{Activity Level} goes up) without saying anything
(\Cc{Announcement Style} \Cc{Silent}) or with just saying or mumbling
something about the issue (\Cc{Announcement Style} \Cb{Implicit}) but never
explicitly announcing the correction action.

Despite the close \Cc{Visual Coupling} (usually 
\Ca{Same Block}, sometimes even \Cc{Same Line}), the \Cb{Driver}
sometimes continues his work (\Cb{Effect on When} is \Cb{Unperturbed}).
He almost never changes his plans due to the intervention
(\Cb{Effect on What} is usually \Cb{Stable}).

Example: The \Cb{Driver} writes \texttt{return x+y*z<t}. 
Without saying anything (\Cc{Announcement Style} \Cc{Silent}), 
the \Cb{Observer} selects the \texttt{y*z} in this line
(\Cc{Visual Coupling} is \Cc{Same Line}) and uses the 
``Extract Local Variable'' dialog to refactor the multiplication 
into a local variable, naming it according to its semantics. 
The \Cb{Driver} recognizes this, just lets it happen (\Cb{Temporizing}),
and then continues with his work (\Cb{Stable}).

\Cc{Direct Fix} is an \refq{FEdit} phenomenon with pleasant properties
(\refq{FPos}).
Its main feature is that it does not disturb the \Cb{Driver}'s 
\Ca{Train of Thought}. 
It also saves (a minor amount of) time.
The activity is short, so the reduction in \Ca{Mental Coupling} appeared to be
unproblematic in our pair. 
It requires trust.
If \Cb{Temporizing} occured, it appeared to provide recreation and to not
interrupt the \Ca{Train of Thought}.
Overall, \Cc{Direct Fix} appeared to lead to higher process
\Cb{Fluency}.

\subsection{Type ``Contribution''}

The \Cb{Observer} will sometimes think of an action that is not (or not yet)
strictly required to proceed but may be helpful when
further pursuing the current \Ca{Train of Thought}.

He will often decide to act alone (with or without announcement) and
similar behavior applies as for \Cc{Direct Fix}.

For example, we have seen the \Cb{Active Observer} visit a requirements document and
copy a few pertinent lines from it into the source code, thus greatly enhancing
the ease of the next few work steps for the \Cb{Driver}.

Another example is a \Cb{Contribution} by looking something up and telling
the \Cb{Driver} about it: 
The \Cb{Observer} realized that the \Cb{Driver} was not
sure about the methods he needed to implement for an interface of the Observer
design pattern \cite{GamHelJoh95}. 
While the \Cb{Driver} stayed in the artifact, the \Cb{Active Observer} visited
Wikipedia \footnote{\url{http://en.wikipedia.org/wiki/Observer_pattern}}
and initiated a short dialog during which the
\Cb{Driver} integrated the information provided by the \Cb{Active Observer}
into the artifact.

This is a behavior that can be of \refq{FView} or of \refq{FEdit} type.
A local pair could emulate it with a quick succession of two driver changes.
Obviously, however, executing such episodes is cumbersome unless there are
two keyboards and two mice.
We have not observed any negative effects, so we count 
\Cb{Contribution} under \refq{FPos}.

\subsection{Type ``Addition''}

Alternatively, the \Cb{Observer} will sometimes think of an action that is 
still related to but beyond the current \Ca{Train of Thought}.
We call such an action an \Cb{Addition}.
For instance, the \Cb{Driver} just wrote a call to a not-yet-existing method
and then created the method by calling the respective IDE function.
The \Cb{Observer} recognizes a property this method will need to
have, tells the \Cb{Driver} he would like to make a note about something, and
then promptly proceeds to make his \Cb{Addition} by adding an appropriate
comment to the method.

Like \Cc{Direct Fix}, \Cb{Addition}, if done well, appears to lead to higher
process 
\Cb{Fluency} by avoiding the negotiation required for a driver change.
If done in an inappropriate manner, it could interrupt the \Cb{Driver}'s 
\Ca{Train of Thought}, but we have not seen indications this ever happened in
our pair, 
so this is another \refq{FEdit} behavior to be counted as \refq{FPos}.

\subsection{Type ``Local Solution''}

Sometimes the next step in a solution process is a moderately-sized, somewhat
self-contained piece of work for which the motivation or expertise of the
\Cb{Observer} is better than the \Cb{Driver}'s.
In local PP and DPP alike, the \Cb{Driver} will then often hand off this piece of
work to the partner.
This behavior, which we call \Cb{Local Solution} is a form of \Cb{Delegation}.
It works much like a subroutine call:
Once the piece of work is finished, the previous work continues where it left
off and typically with the previous \Cb{Driver}.
For instance our pair was at a point where they needed so sort a list of files,
so they needed a file comparator object. 
Without any discussion, the former \Cb{Observer} took the \Cb{Driver} role
and searched the web for an example implementation of a file comparator, 
pasted it into the source code, adapted it it for the current needs, and 
gave back the control to the previous \Cb{Driver} who continues his work.

The crucial point about this \refq{FEdit} (or possibly sometimes
\refq{FView}?) behavior appears to be not so much the execution of the driver
change -- the episode is long enough to amortize its effort.
The point appears to be that the motivated \Cb{Observer} can jump right into
action, without any need to negotiate about it.
A \Cb{Local Solution} episode allows the original \Cb{Driver} to 
\Cb{Implicitly Step Back} from his role \emph{after} the partner has taken it
(and is apparently doing something sensible) rather than making an explicit
decision to do so.

Conflicts about \Cb{Local Solution} behavior or negative consequences
resulting from it are easily conceivable, but we found none for our pair,
so this is another behavior pattern that can be \refq{FPos} without
carrying \refq{FNeg} along and again higher process \Cb{Fluency} is the right
way to describe the result.

\subsection{Type ``Co-Driving''}

When both partners intensely and equally contribute to the solution process,
the difference between \Cb{Driver} and \Cb{Active Observer} disappears; 
the two are \Cb{Co-Driving}, thus introducing another variant of the classical
driver/observer roles. 
Such collaboration can be described as a having maximum 
\Ca{Mental Coupling}: Both partners have an identical and very good grasp of
what needs to be achieved and work together intensely and closely on finding
out how and doing it.
Their actions are interleaved on a fine grain and there is no discrimination
of the \Cb{Activity Level} or the types of the actions any more between the
two.

We observed \Cb{Co-Driving} for instance during the difficult creation of
a complex boolean expression and when coding the production of an
expressive error message involving multiple method calls.

\Cb{Co-Driving} is an \refq{FEdit} behavior for which eDPP really excels and
obviously beats even double-keyboard local PP; 
it is again about \Cb{Fluency} and clearly belongs to \refq{FPos}.

\subsection{Type ``Parallelization''}
The pair explicitely splits off to work in parallel on different aspects of
the same
\Ca{Train of Thought} in the \Cc{Same Viewport}. 
This can be considered a two-sided \Cb{Local Solution}: Each partner
delegates something to the other and both work concurrently.

As in \ref{awarenessdialogs}, the split-off in our pair was preceded by an
explicit agreement regarding the concrete sub-task each one has to do, e.g.:\\
Dev: \Qu{Could you just adapt the thing above, the regex?}\\
Arch: \Qu{Sure.} \\
Dev: \Qu{And meanwhile I'll write the calendar here.} \\
Arch: \Qu{Yeah, do that.}

This strong \refq{FEdit} behavior can be seen as switching to side-by-side
programming, but on a very fine-grained level that allows both partners to
stick with the same overall \Ca{Train of Thought}. 
The partners split off with well-defined and hardly demanding
sub-tasks.
This mode might allow to escape the possible inefficiency of PP for some
combinations of simple sub-tasks and would then belong into \refq{FPos}, 
but it obviously also runs the risk of inadvertantly leaving pair programming
so may also count as \refq{FNeg}.
We have not seen enough cases of \Cb{Parallelization} to judge 
the pros and cons even for our one pair.

\section{Limitations and Threats to Validity}
\label{limitations}

The pair we observed were strong software developers with 
good communication skills and were very familiar with each other. 
We conjecture these are necessary conditions for the degree of eDPP success
described above.
We do not know what might be sufficient conditions nor do we now how common
the above conditions are, so we can say very little about the
generalizability of our results.
(This is not a big problem, because the existence proof is interesting
enough.) 

The amount of material (in terms of the number of hours as well as variety of
pairs) we have studied is still small, so the list of phenomena we report is
likely incomplete. 

The conceptualization is so far fairly local; the episodes
described are short and do not yet represent substantial solution processes
completely. 
This means the overall role that the phenomena described play in the
pair process as a whole is hardly understood so far.

At least for editing, our report emphasizes phenomena that together cover
only a small fraction of the time of the overall pair programming session; 
one should not overrate their overall importance.
(This is also not a big problem, because the absence of problems is benefit
enough.) 

It is conceivable that differences exist between local PP and eDPP
outside the realms of awareness and editing freedom.
Our study was not designed to detect such differences.

Note that the limitations of the Saros tool do \emph{not} add to the
limitations of the study, because if an imperfect tool can still produce eDPP
success, that is all the better.

\section{Conclusions}
\label{conclusion}

\subsection{Regarding awareness}

Conventional wisdom suggests the reduced physical awareness and limited
workspace awareness of a Saros-based eDPP
situation will be a big obstacle against successful collaboration, so we
expected to find many resulting problems, but we have in fact seen almost none.

We conjecture that somehow program code is a great basis for focused and
successful communication and skilled programmers (at least if they are
familiar with each other) have much lower needs for physical awareness than
previously assumed.
As a result, eDPP can work well.

\subsection{Regarding Editing Freedom}

Our pair had the strong and clear intention of performing pair programming,
because besides the practical work results knowledge transfer was an
important goal.

Nevertheless, and although the resulting work process clearly \emph{was} 
pair programming overall in terms of the closeness of collaboration,
the pair \emph{did} make use of the concurrent editing capabilities of eDPP.

We conclude that eDPP (or, as the title calls it ``distributed-pair
programming'') should not be confused with simply distributed PP (or, as the
title calls it, ``distributed pair-programming'').

Furthermore, and most importantly, we found that our pair used concurrent
editing wisely, in a very limited fashion, resulting in several
interesting behaviors we call 
Direct Fix, 
Contribution, 
Addition,
Local Solution,
Co-Driving,
and
Parallelization.
These behaviors created no observable disadvantage but a number of small
advantages that can be summarized as leading to a (slightly) improved fluency
of the work process.

We conjecture that the previous
joint work experience and resulting familiarity and trust of our pair as
well as good communication skills are important preconditions for such eDPP
success.
Given the preconditions (whatever they really are), Saros appears to be a
strong-enough tool to support industrial eDPP successfully.

\subsection*{Acknowledgments}
This work was partially supported by a DFG grant.
We thank J1 and J2 for allowing us to record and scrutinize their sessions.

\bibliographystyle{abbrv}
\bibliography{schenk,general,dpp,agse}


\end{document}